# Coupling hBN quantum emitters to 1D photonic crystal cavities

*Johannes E. Fröch, Sejeong Kim[*], Noah Mendelson, Mehran Kianinia, Milos Toth, Igor Aharonovich[*]*

School of Mathematical and Physical Sciences, University of Technology Sydney, New South Wales 2007, Australia

[*]Sejeong.Kim-1@uts.edu.au, igor.aharonovich@uts.edu.au

**Abstract**

**Quantum photonics technologies require a scalable approach for integration of non-classical light sources with photonic resonators to achieve strong light confinement and enhancement of quantum light emission. Point defects from hexagonal Boron Nitride (hBN) are amongst the front runners for single photon sources due to their ultra bright emission, however, coupling of hBN defects to photonic crystal cavities has so far remained elusive. Here we demonstrate on-chip integration of hBN quantum emitters with photonic crystal cavities**

**from silicon nitride (Si$_3$N$_4$) and achieve experimentally measured Q-factor of 3,300 for hBN/Si$_3$N$_4$ hybrid cavities. We observed 9-fold photoluminescence enhancement of a hBN single photon emission at room temperature. Our work paves the way towards hybrid integrated quantum photonics with hBN, and outlines an excellent path for further development of cavity quantum electrodynamic experiments and on-chip integration of 2D materials.**

Quantum information processing applications require integration of non-classical photon sources with optical cavities and resonators to achieve efficient interfaces between flying qubits and quantum memories as well as for quantum repeater applications.[1-6] Several solid-state systems have been explored to achieve this goal, including III-V quantum dots, color centres in diamond and single molecules.[7-15] Often, there is a trade-off between the optical properties of the selected source in terms of brightness, stability, or tunability and the selection of the cavity material in terms of ease of fabrication and the ability to achieve strong light confinement. The ultimate goal is to engineer a system in which a bright solid-state emitter can be coupled to a scalable on-chip optical resonator and a waveguide, to realise enhanced emission via the Purcell effect.[16-21] Such a system will then become a key building block for cavity quantum electrodynamics experiments and integrated quantum photonic circuitry.

Particularly promising sources are quantum emitters in few-layer hexagonal boron nitride (hBN).[22-30] Due to both high quantum efficiencies and low Debye-Waller factors, they are amongst the brightest solid-state emitters studied so far.[31] Furthermore, the chemical and thermal robustness of hBN as a host material provides an ideal platform free from any severe degradation over long time periods.[32, 33] A step forward in the realization of practical devices with quantum emitters in hBN, is to couple them to photonic crystal cavities (PCCs), which alter the local photonic density of states leading to a decreased linewidth and an increase in their emission rates, known as Purcell enhancement. However, to date, integrating hBN quantum emitters with on-chip photonic cavities has been a challenging task,[34, 35] mostly due to a broad and random distribution of their emission wavelengths in the visible range, which has made coupling of the emitters to PCCs inherently impractical.

Here, we report the on-chip integration of hBN quantum emitters with one-dimensional photonic crystal nanobeam cavities fabricated from silicon nitride ($Si_3N_4$). Specifically, we take advantage of the recently established method to grow thin hBN layers with a unimodal distribution of quantum emission at (580 ± 10) nm and spatial densities as high as 2 µm$^{-2}$.[36] Silicon nitride is used in this study not only because it is available on a large scale at a reasonable cost, but also because the refractive index contrast between $Si_3N_4$ (n=2.0) and hBN (n=2.1) is negligible. The established fabrication protocols enable us to benefit from high Q-factor (Q) photonic cavities which corresponds to 3,300. We were able to achieve coupling of several quantum emitters to PCCs with a photoluminescence (PL) enhancement up to 9 times. Our results outline a promising route towards scalable on-chip hybrid photonic networks that harness the properties of quantum emitters in hBN as quantum light sources and $Si_3N_4$ as a technologically-mature material platform.

**Results/Discussion**

Here, it is important to emphasize the non-trivial methodology we employed to create the hybrid quantum photonic devices. Traditionally, in the hybrid photonic approach,[37] the photonic components are fabricated first, and the target material which requires enhanced light interaction is then transferred either using a pick-and-place method (e.g. for nanodiamonds)[38], or using a wet transfer method (e.g. for 2D materials)[39-43]. Such a sequence often results in breaking of freestanding resonators[40] or degrading of their functionality due to contamination or imperfect spatial alignment. In our approach, we reverse the sequence, and first transfer the hBN onto the $Si_3N_4$, before fabricating the final devices, thus eliminating the above mentioned technical difficulties.

A step-by-step scheme of the device fabrication is depicted in Figure 1a, whilst specific conditions, such as the used equipment are detailed in the Methods section. First, (I) a few-layer hBN film was grown on copper utilizing a Chemical Vapor Deposition (CVD) method, yielding quantum emitters with a unimodal distribution of the zero phonon line (ZPL) centered at 580 nm and a spatial density of ~ 2 $\mu m^{-2}$. To ensure the quality of the hBN film before further processing, (II) a pre-check was performed to confirm the presence of quantum emitters in the as-grown material (Supplementary Information 1). Then, (III) the hBN film was transferred from copper to the $Si_3N_4$ substrate (180 nm on Si) using a wet transfer technique, followed by an annealing step (1 Torr Ar, 850 °C, 30 min). An annealing step at this stage was required in order to avoid delineation of the transferred film in the later steps (Supplementary Information 2). Subsequently, (IV) a polymer resist film (CSAR 62) was spin-coated to a thickness of approximately 500 nm and (V) the photonic cavity design was patterned using electron beam lithography (EBL) and developed. In the current work, we chose to work with 1D PCCs as they provide small mode volumes combined with high-Q resonances. Finally, (VI) the resist was then directly employed as a hard mask to transfer the pattern into the underlying hBN/$Si_3N_4$ film by reactive ion etching (RIE) in $SF_6$. After RIE, the resist was stripped in warm acetone and the device structures were released from the underlying silicon using a KOH wet etch. These fabrication steps are in principle

CMOS compatible and with recent reports of direct growth of hBN on $Si_3N_4$,[44] may lead to a transfer-free device fabrication process in the future.

An optical microscope image of the fabricated sample is shown in Figure 1b. The hBN film is directly discernible from the substrate by optical contrast with the boundaries outlined by white dashed lines as a guide to the eye, still adhering well to the $Si_3N_4$ substrate without folding of the film over the device area. A representative SEM image of such an array is shown in Figure 1c.

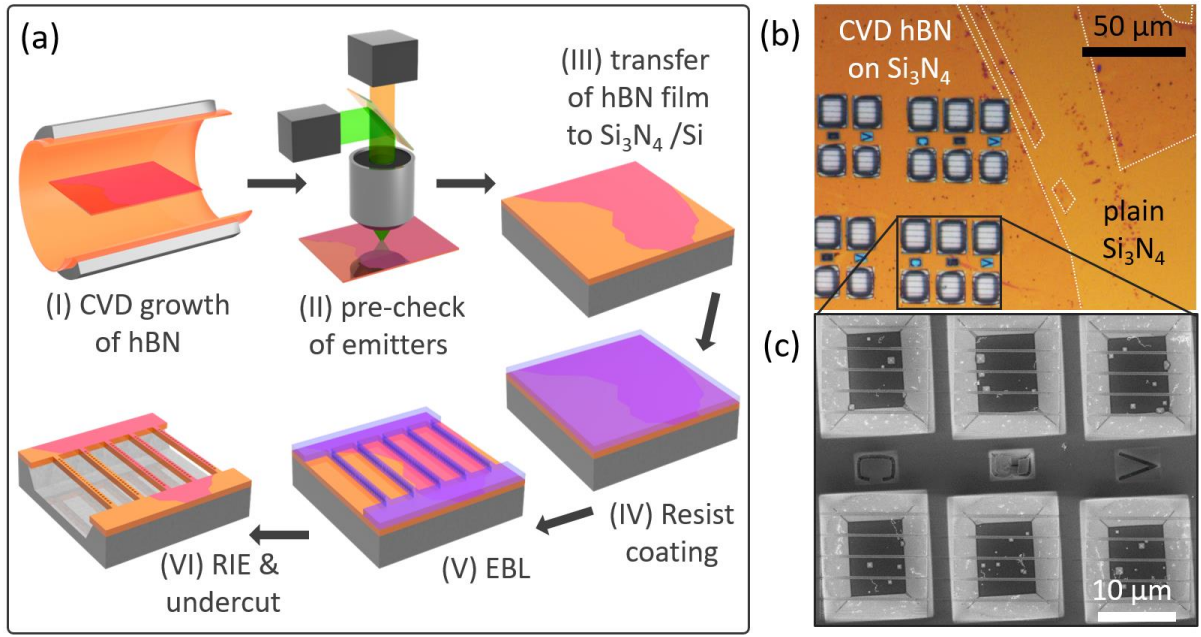

*Figure 1. Integration of CVD-grown hBN with 1D photonic crystal cavities. (a) Schematic of the fabrication steps: (I) CVD growth of hBN on copper. (II) Pre-check of the hBN film on copper using a confocal microscope setup. (III) Transfer the hBN from copper to a 180 nm thick $Si_3N_4$ substrate. (IV) Spin coating of EBL resist (CSAR 62). (V) EBL patterning of the cavity structures. (VI) RIE and undercutting to create free-standing hBN/$Si_3N_4$ devices. (b) An optical microscope image with the final sample. hBN-covered area (white-dashed guideline) is directly visible due to optical contrast. (c) An SEM image showing the final free-standing, hybrid photonic cavity arrays.*

An image of a final device comprised of a hBN film integrated with a 1D PCC is shown in Figure 2a. Note that no 2D material is draping over the device as the hBN film is integrated within this architecture. The employed design is a 1D photonic crystal nanobeam with width (*w*) of 300 nm, thickness (*h*) of 180 nm, periodicity (*a*) of 230 nm, and airhole radius (*r*) of 80 nm.[45] A photonic cavity is introduced by linearly tapering 10 airholes both in radius and periodicity, for which the center of the cavity has a periodicity of $0.85a$ and airhole radius of $0.75r$, respectively. The photonic device is analysed based on the finite-difference time-domain (FDTD) method using a commercial photonic simulation tool (Lumerical Inc.). We assume a 5 nm-thick hBN layer ($n_{hBN}$=2.1) on top of 180 nm $Si_3N_4$ ($n_{SiN}$=2.0). The electric field intensity profile in x-y plane and x-z plane are depicted in Figure 2b and 2c, respectively. Here the intensity profile in Figure 2b is extracted from the center of the hBN film. The theoretical Q-factor and mode volume $V_m$ ($V_m = \int \varepsilon E^2 dV / \max(\varepsilon E^2)$) at a resonant wavelength of 590 nm correspond to 16,000 and ~ 1.5 $(n/\lambda)^3$, respectively. The spontaneous emission rate enhancement of an emitter in the hBN film is given by the Purcell factor *F*, defined in Equation (1). In parts the enhancement is determined by the Q-factor and mode volume ($V_m$), characteristic values of the cavity. Whilst the field intensity ($E^2(r)$) at the emitter location (*r*) and the angular mismatch (*ξ*) between emission dipole and cavity polarization depend on the position and orientation of the emitter.

$$F = \frac{3}{4\pi^2}\left(\frac{\lambda}{n}\right)^3 \left(\frac{Q}{V_m}\right) E^2(r_E) \cos^2(\xi) \quad (1)$$

In addition to the advantage of the CVD-grown hBN that alleviates issues of spectral matching, we also achieve a considerably high maximum field intensity at the emitter location, approaching 40 %, as shown in the top view of Figure 2b. Because of the direct integration method the optical mode is confined within the hBN film. This is advantageous compared to previous works that employed the evanescent field of the mode to couple to emitters.

To directly demonstrate the influence of the hBN film on device functionality we compared adjacent regions comprised of bare (blue outline) and hBN-covered (red outline) cavities as shown in Figure 2d, respectively. Device characterization was done in a lab-built confocal microscope setup with a 500 µW, 532 nm excitation source, exciting and collecting PL through a 0.9 NA objective. The setup is described in the Methods section and schematically depicted in Supplementary Information 3. Representative cavity modes are in the range from 580 nm to 605 nm, which are displayed to the right side of the respective areas. All measured Q-factors ($Q=\lambda/\Delta\lambda$) from this region are plotted in Figure 2e, which illustrates that the performance of cavities that were covered with hBN (2,700 ± 530) is on par with plain cavities of (2,730 ± 430). The almost identical device performance emphasizes the advantage of the developed fabrication protocol that was used in this study.

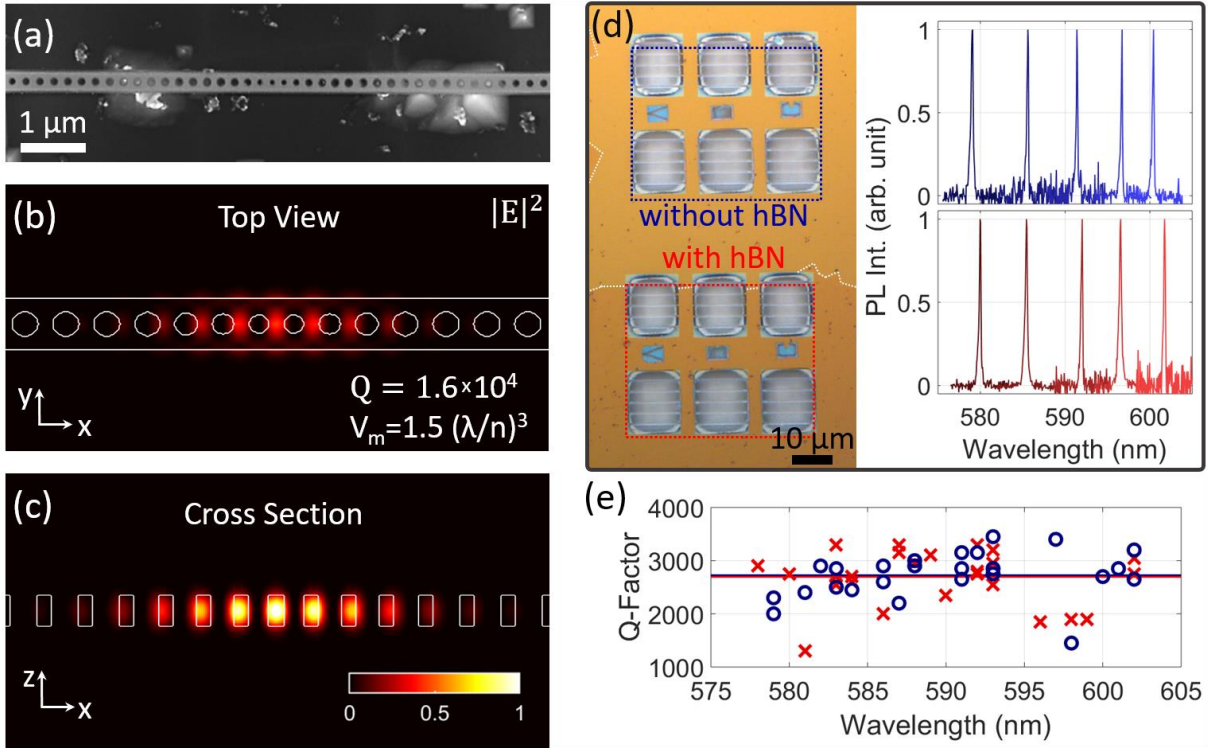

*Figure 2.* Photonic cavity design and properties. (a) Magnified view of a fabricated 1D nanobeam photonic crystal cavity with circular holes. Towards the middle of the cavity, the hole periodicity and diameter are linearly tapered. (b) Top view FDTD simulation showing the main resonance at 590 nm in the center of the device with a theoretical Q-factor of $1.6\times10^4$ and a mode volume of $1.5\ (n/\lambda)^3$. (c) Side view showing the distribution of the mode and the overlap with the top hBN layer, for which the field overlap is approximately 40% for a 5 nm thick hBN film. (d) Optical image of an area of fabricated devices at the boundary of a transferred hBN film. A direct comparison of the device functionality was taken from cavities fabricated under the same conditions without hBN and cavities with hBN on top showing representative resonances at a target wavelength spanning from 580 nm to 605 nm shown to their right, respectively. (e) A comparison of the Q-factor from the cavities in the shown area reveals on average values of (2,730 ± 430), and (2,700 ± 530) for cavities without (crosses) and with hBN (circles), respectively.

We observed quantum emitters at the center of PCCs, whose ZPL partially or highly overlapped with a resonant mode. Characterization of emitter-cavity systems was done at room temperature under 532 nm laser excitation (cw, 75 µW). Figure 3a and 3b are examples of two coupled emitters, which are referred to below as Cavity 1 and Cavity 2, respectively. For both cases we observed the emitter location to be near the nanobeam center, as evidenced by PL maps in Figures 3a and 3b (I). Spectra (II) collected from the white outlined spots show ZPLs at 588 nm and 593 nm, respectively (red curves). Whilst for Cavity 1 the spectral features of the emitter are distinctive from the cavity resonance at 595 nm, due to a low degree of coupling, for Cavity 2 the shape of the ZPL could not be clearly distinguished from the cavity mode due to a high degree of coupling. Q-factors of both cavities (insets) were determined by a Lorentzian fit, which correspond to 1,700 and 2,400, respectively. To confirm their single photon nature, we spectrally filtered the collection signal by band pass filters (578 - 598 nm, and 583 - 603 nm, respectively) and measured the second-order correlation functions, $g^{(2)}(\tau)$. Experimental data is fitted (without background correction) to a double exponential function as shown in red, and we obtained $g^{(2)}(0)$ values of 0.3 and 0.1 for the two emitters.

To estimate the PL enhancement at the cavity resonance, we further fitted the respective ZPL and cavity mode as a sum of Lorentzian functions (purple), shown in (IV). For Cavity 1, we considered a sum of three Lorentzians corresponding to the main ZPL (red), the broader red tail of the ZPL at lower energies (yellow), and the cavity mode (blue). We note here that the role of the 2$^{nd}$ Lorentzian is still not fully understood and current point of debate in the literature, assuming it to be a 2$^{nd}$ ZPL or an acoustic phonon sideband.[46] In the case of Cavity 2, a sum of two Lorentzians was used for the fit, assuming that the main portion of the coupled system is a Lorentzian centered at 593 nm (blue) and a partially uncoupled component (red) centered at 592.5 nm. To evaluate the PL enhancement due to the Purcell effect, we compare the relative intensities of the cavity mode extracted from the fits in Figure 3 (IV) relative to the uncoupled emission. In particular, we calculate the enhancement factor as the total emission at the cavity resonance (purple) over the total emission minus the cavity related emission (blue). Thus, Cavity 1 exhibits a PL enhancement of 1.5, and Cavity 2 is estimated to have 9 times PL enhancement. This PL enhancement is the highest achieved value for hBN quantum emitters to date. It can be further increased by improving the field overlap with the resonant mode, e.g. by increasing the angular match between the emitter dipole orientation and the cavity polarization or increasing the spatial match between the emitter location and the mode intensity maximum. Purcell factors can be ideally extracted by measuring lifetime reduction at cryogenic temperatures, i.e. when the emitter linewidth narrows significantly to the order of the cavity mode. Unfortunately, at this point most studied hBN quantum emitters bleached within the time frame of the experiments, which prohibited controlled tuning of the same system. Techniques to mitigate this problem are currently being pursued.

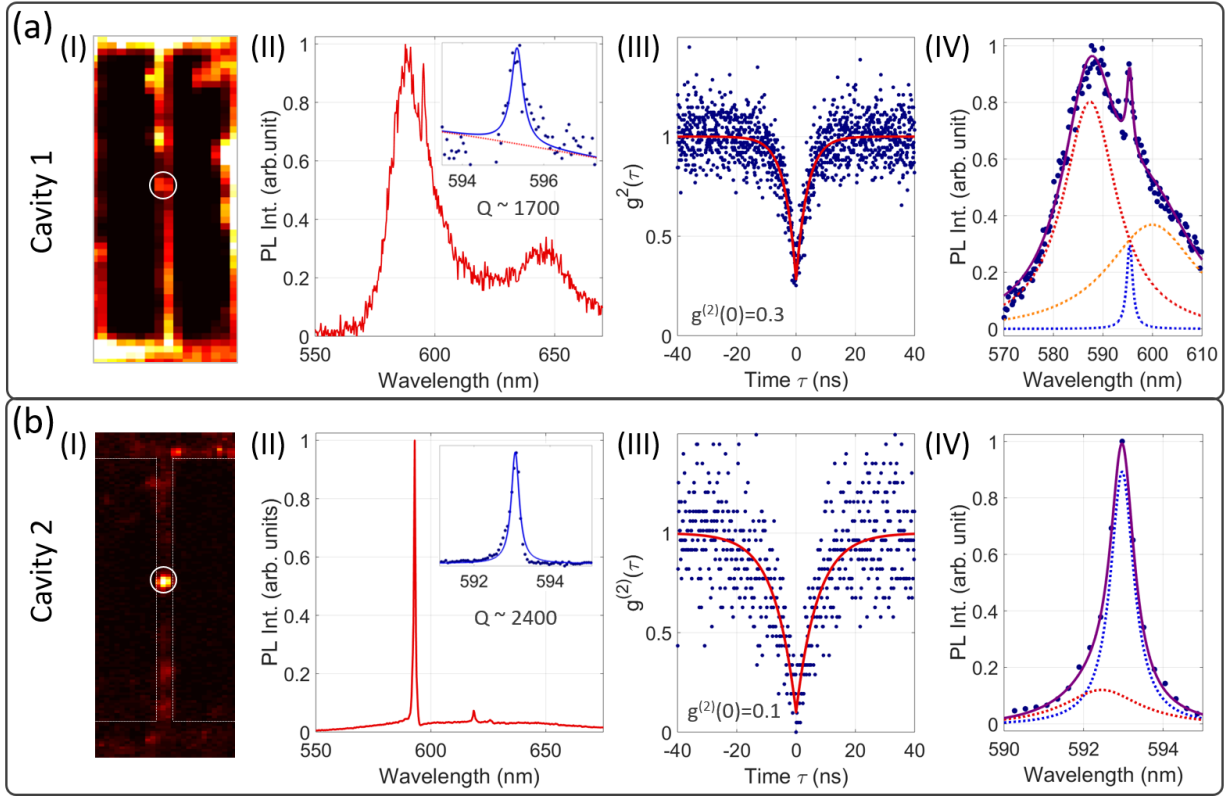

*Figure 3.* Coupled hBN - Si$_3$N$_4$ system. Cavity 1 (a) shows partial coupling, whilst Cavity 2 (b) shows a high degree of coupling. (I) PL maps show the emitter positions within the respective cavities, being at or close to the center of the device. (II) Spectra from the respective white-marked spots show the emissions of quantum emitters at 588 nm and 593 nm for Cavity 1 and Cavity 2, respectively. For Cavity 1, a distinctive resonance mode can be seen at 595 nm, whilst for Cavity 2 the ZPL cannot be clearly distinguished from a cavity mode. Insets show the cavity resonance determined from a measurement with a higher resolution grating. (III) Correlation measurements with $g^2(0)$ values of 0.3 and 0.1, demonstrating the single photon nature of the emitters. (IV) The ZPL from the coupled system was, in each case, fitted with a sum of Lorentzians (purple) to estimate the respective contributions from emitter (red and yellow) and cavity (blue).

Nonetheless, occasional spectral jumps of the emissions enabled us to directly observe coupling between quantum emitters and cavity modes. For hBN quantum emitters, spectral diffusion over a range of several nanometers is common due to fluctuations in local electric fields caused by charge (de)trapping at surrounding defects.[47] Figure 4 shows an example of a spectrally jumping quantum emitter and a fixed cavity resonance. Time series of normalized PL spectra to the maximum intensity of the cavity mode are plotted in Figure 4a to clearly show the shift of the emitter (blue arrow). The ZPL of a quantum emitter was initially identified at 598 nm, close to a cavity mode at 593 nm. Whilst it was only partially overlapping at first, we observe that during the course of the PL measurement, a 1$^{st}$ shift of the ZPL occurred, which resulted in an overlap of the ZPL and the main cavity mode. Consequently, a 2$^{nd}$ shift of the ZPL detuned the emitter from the cavity mode to 584 nm, followed by photobleaching, whilst the cavity mode was still observable. To determine the PL enhancement, the same spectra are displayed without normalization in Figure 4b, showing that the intensity at the ZPL increased from 1.9 to 9.3 after the 1$^{st}$ shift, and decreased to 1.5 after the 2$^{nd}$ shift, thus we estimate the PL enhancement within the range of 5 – 6. At the same time Lorentzian fits in Figure 4c show that for spectra 1 and 3 the main portion of the ZPL (red) can be clearly distinguished from the cavity mode (blue), whilst for spectra 2 the main contribution to the ZPL is virtually inextricable from the resonance and only a second phonon related part (yellow) contributes to the spectrum. Therefore, as the emitter location is fixed and the dipole orientation change is negligible, the PL enhancement in the second spectra is due to the maximum spectral matching between the optical mode and the quantum emitter.

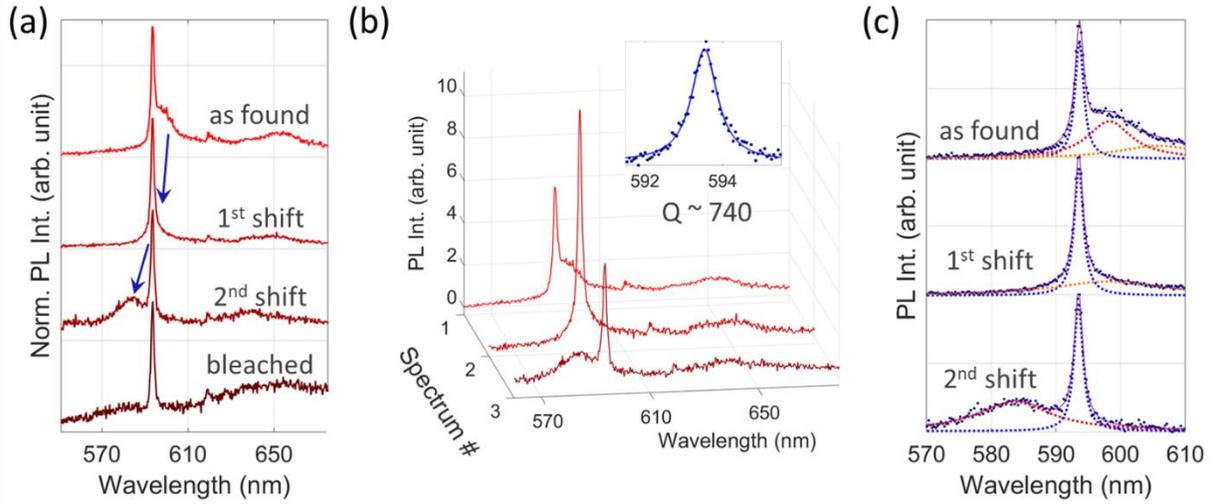

*Figure 4. Dynamic enhancement of a quantum emitter which spectrally diffused to a cavity mode. (a) PL spectra, which are normalized to the maximum intensity of cavity mode, measured during spectral jumping of the emitter. As found, the emitter was partially overlapping with a cavity mode at 593 nm. A 1st spectral shift caused the ZPL to be coupled with the cavity mode. After a 2nd shift, the ZPL was again detuned by 5 nm from the resonance. Then we observed bleaching of the quantum emitter. (b) The same PL spectra without normalization showing PL enhancement of the emitter and an inset of the Q-factor. (c) Lorentzian fits of spectra in (a). The ZPL (red and yellow) and the cavity mode (blue) were used to estimate the relative contributions, shown in the same order as in (a).*

**Conclusion**

To conclude, we demonstrated a scalable approach for on-chip integration of quantum emitters in hBN with $Si_3N_4$. We overcome previous challenges where the ZPL was occurring randomly within the visible spectrum, prohibiting systematic investigations on coupling with dielectric cavity resonances. In particular, we make use of few-layer hBN grown by a recently developed method, which yields a clear unimodal distribution of the ZPL centered at 580 nm. Integration prior to the fabrication process allowed us to achieve large uniform device areas, and to maintain high Q-factors up to about 3,300. By direct comparison with devices from plain $Si_3N_4$ we have shown that the functionality of devices integrated with hBN and without hBN are essentially unchanged, underlining the advantage of this processing approach. Owing to the high spatial density of emitters, we observed several cases of quantum emitters coupled to the mode resonance, achieving 9-fold emission enhancement.

At this point further experimental work was limited by bleaching of the quantum emitters. Therefore, while our work lays solid foundation towards the realisation of integrated quantum photonics with hBN, efforts must be concentrated to the improvement of the stability of hBN quantum emitters. This could potentially be achieved at the same time by encapsulating and protecting the hBN layer with a top $Si_3N_4$ film. Such an approach would make large scale nanophotonic chip fabrication feasible. Ultimately, the devised approach outlines the path forward to more sophisticated experiments using such hybrid systems based on hBN quantum emitters for integrated quantum networks with potential applications for quantum memories and applications as quantum repeaters.

**Acknowledgements**

The authors thank the Australian Research Council (DE180100070, DE180100810, DP190101058) and the Office of Naval Research Global under grant number N62909-18-1-2025 for financial support.

**Methods**

**Device fabrication**

$Si_3N_4$ substrates were prepared by purchasing wafers of 300 nm thick LPCVD $Si_3N_4$ (MTI Corporation), which was then slowly etched using boiling $H_3PO_4$ (85 wt. % ~ 180 °C) until the desired thickness was achieved, determined by the visual contrast change. The growing method of the few layered hBN film was detailed in our previous work[36]. For wet transfer the hBN/ copper samples were spincoated with ~ 200 nm PMMA A3 (Microchem.) and baked at 90 °C for 3 min. The supporting copper was then etched in a 0.5 M $(NH_4)_2S_2O_8$ solution until the PMMA/ hBN film delineated. The floating film was picked up with a clean Si substrate and the film was transferred to Ultrapure Water (UPW) 3 times to remove residual contaminants. The target $Si_3N_4$ substrate was then used to pick up the cleansed floating film and placed on a hot plate at 100 °C for 10 min to remove residual liquids. The PMMA was then stripped in a warm Acetone bath overnight and briefly (5 min) placed in a UV Ozone cleaner to remove residual contaminants. In a following step the sample was annealed in a Tube Furnace (Lindberg Blue Mini-mite) at 850 °C for 30 min under continuous Argon Flow at a pressure of 1 Torr, to increase the adhesion to the substrate and further eliminate contamination. The resist for EBL was AR-P 6200.11 (AllResist GmBH.), which was spin coated at 3500 rpm for 1 min to give an approximate thickness of ~ 500 nm. EBL was conducted in a FEG-SEM (Zeiss Supra 55 VP) equipped with a RAITH EBL system. The parameters used for patterning were 20 pA at 30 kV and base dose of ~ 100 $\mu C\ cm^{-2}$. The patterns were developed in AR 600-546 (AllResist GmBH.) for ~ 45 s at a Temp of ~ 5 °C. After development the sample was placed in a UV Ozone cleaner for ~ 30 s to descum the developed regions. Pattern transfer was then conducted in a self-biased Reactive Ion Etching System using a (~ 300 V self-bias) and a Forward Power of ~ 90 W in a $SF_6$ Plasma at a Flow rate of 60 SCCM and pressure of ~ 10 mTorr. After etching the resist was stripped in a warm Acetone bath overnight, whilst remaining resist was further removed by calcination on a hot plate in air at 400 °C (~ 2hrs) and a UV ozone treatment (~ 10 min.). Cavity structures were then released by etching the underlying Silicon in a warm KOH (20 wt. %) solution for ~ 20 mins. Following the KOH etch the substrate was put in 3 consecutive lukewarm UPW baths to remove KOH residues (15 min ea.). After the UPW baths the substrate was put into IPA for the structures to be immersed in a liquid with a lower surface tension, minimizing the stress during the final drying under a gentle blow of $N_2$. After all wet processing steps the substrate is once more cleaned in a UV ozone step for ~ 10 minutes.

**Simulations**

All numerical modelling including calculation of Q-factors, mode volumes and Field Intensity maps was performed by using the finite-difference time-domain (FDTD) method. The computational domain was defined as a cuboid volume of 12×1.3×1.3 μm$^3$ with a minimum mesh size of 5 nm. The refractive index for silicon nitride used is 2.0.

**PL characterization**

Optical characterization was done in a lab-built confocal setup, which is schematically depicted in the Supplementary Information 3. For characterization of cavities and identification of quantum emitters a 532 nm laser (GemStone) is collimated, attenuated, and guided by several dielectric mirrors and a dichroic mirror (Semrock Di03-R532-t1) to a scanning mirror, which reflects the laser to a 4*f* system (2 achromatic doublets, f= 150 mm) into a 0.9 NA objective focusing onto the sample forming a diffraction limited excitation spot. Luminescence is collected by the same objective, guided back and transmitted through the dichroic mirror. A longpass filter (Thorlabs 550 FEL) filtered any residual laser leakage before coupling the PL either to a fiber connected to a spectrometer (Princeton Instruments) or a fiber based Hanburry-Brown and Twiss (HBT) setup connected to 2 APDs (Excelitas). For characterization mentioned in Figure 2, the laser power was attenuated to 500 µW before the objective. For identification of quantum emitter the power was attenuated to ~ 75 µW. For spectral characterisation, PL of quantum emitters were dispersed via a 300 l/mm grating and collected for 20 s (if not mentioned otherwise), whilst for characterization of photonic crystal cavities the grating was set to 1200 l/mm and collected for 40 s. Time correlation experiments were done leading the APD signal to a PicoHarp 300 Picoquant and measurements were not corrected for background.